\documentclass[letterpaper]{article}
\usepackage{aaai2026}
\usepackage{times}
\usepackage{helvet}
\usepackage{courier}
\usepackage[hyphens]{url}
\usepackage{graphicx}
\usepackage{natbib}
\usepackage{caption}
\usepackage{booktabs}
\usepackage{multirow}
\usepackage[table]{xcolor}
\usepackage[accsupp]{axessibility}
\usepackage{amsmath}
\nocopyright
\title{Safeguarding Text-to-Image Generative Models Against Unauthorized Knowledge Distillation}
\author{
    Yilan Gao\textsuperscript{\rm 1},
    Sida Huang\textsuperscript{\rm 1,2},
    Hongyuan Zhang\textsuperscript{\rm 3}\thanks{Corresponding author.},
    Xuelong Li\textsuperscript{\rm 2}\footnotemark[1]
}
\affiliations{
    \textsuperscript{\rm 1}School of Artificial Intelligence, OPtics and ElectroNics (iOPEN), \\Northwestern Polytechnical University, Xi'an 710072, P.~R.~China\\
    \textsuperscript{\rm 2}Institute of Artificial Intelligence (TeleAI), China Telecom, P.~R.~China\\
    \textsuperscript{\rm 3}The University of Hong Kong\\
    2022yl@mail.nwpu.edu.cn, sidahuang2001@gmail.com, hyzhang98@gmail.com, xuelong\_li@ieee.org
}

\begin{document}

\maketitle

\begin{abstract}
Closed-weight generative services are increasingly deployed through query-based APIs, where users can obtain generated outputs while model parameters remain inaccessible. However, such deployment does not prevent model stealing: an attacker can repeatedly query the service, collect large volumes of released synthetic images, and use them as training data for a private substitute model. This query-output-driven process enables unauthorized knowledge distillation and capability replication without direct access to the original weights. To mitigate this threat, a practical defense should preserve the visual fidelity of released images, provide explicit control over perturbation magnitude, and scale efficiently to large-volume output release. We present \textbf{WaveGuard}, a single-pass, generator-based protection framework that safeguards released synthetic images under a user-specified perturbation budget. WaveGuard employs a frequency-aware perturbation generator to inject structured, imperceptible perturbations that maintain perceptual utility for benign viewers while reducing the usefulness of protected images as training data for unauthorized student models. Extensive experiments under WikiArt-related synthetic-output distillation settings show that WaveGuard achieves a favorable efficacy--fidelity--efficiency trade-off, with explicit imperceptibility control and substantial gains in protection efficiency.
\end{abstract}

\section{Introduction}\label{sec}

Generative models~\cite{song_denoising_2022,podell_sdxl_2023,huang_nfig_2025,gu_rectified_noise_2026} have become powerful tools for image synthesis~\cite{huang_laytrol_2026}, personalization, and creative content generation~\cite{zhu_viewmask_2025}. In practical deployments, model owners often expose these systems through closed-weight services: users can query the service and obtain generated outputs, while the model parameters remain private. However, keeping the weights inaccessible does not eliminate the risk of model stealing~\cite{ma2021undistillable}. Since generated images themselves encode the visual knowledge, stylistic patterns, and generation capability of the service, an attacker can repeatedly query the model, collect a large number of released outputs, and use them as training data for a private substitute model.

This threat differs from conventional model stealing scenarios that mainly exploit prediction labels, logits, or decision boundaries. For generative models, the released samples can directly serve as supervision for downstream training. In this setting, unauthorized distillation does not require access to the teacher model's weights, training data, or internal representations; instead, the attacker distills knowledge from the synthetic outputs exposed by the service. The resulting substitute model may imitate the teacher's visual style or generation behavior, causing intellectual property infringement and weakening the protection offered by closed-weight deployment. In this work, we focus on image generative services and study this threat as \textbf{unauthorized knowledge distillation from released synthetic images}.

\begin{figure}[tb]
    \centering
    \includegraphics[width=\linewidth]{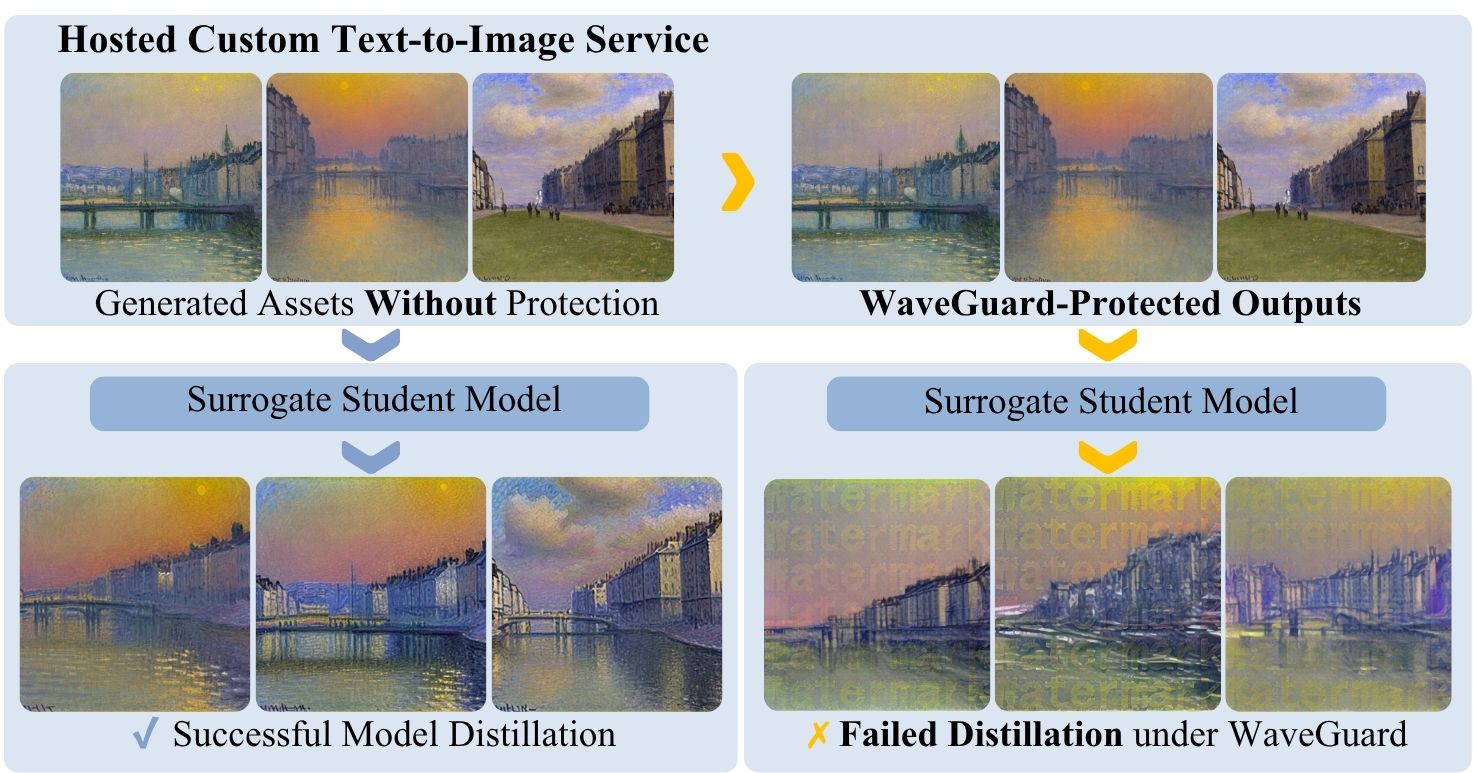}
    \caption{
    \textbf{Unauthorized distillation from released synthetic images and WaveGuard protection.}
    (Left) An attacker queries a closed-weight generative service, collects released synthetic images, and trains a substitute model to imitate the teacher.
    (Right) WaveGuard applies bounded, structured perturbations before release, reducing downstream imitation while preserving visual fidelity.
    }
    \label{fig}
\end{figure}

A practical defense for this setting must satisfy several deployment-oriented requirements. First, it should reduce the usefulness of released images for unauthorized substitute training. Second, it should preserve the visual fidelity of the released outputs, since these images are still intended for benign users or public presentation. Third, it should provide explicit control over perturbation magnitude, so that the protection strength can be adjusted according to a user-specified imperceptibility budget. Finally, it should scale to large-volume output release, where expensive per-image optimization becomes impractical.

Existing protection methods only partially satisfy these requirements. Watermarking-oriented methods~\cite{cui_diffusionshield_2024,zhao_recipe_2023} are useful for attribution, ownership verification, or post-hoc tracing, but they do not directly prevent released images from being reused as training data. Diffusion-oriented anti-personalization methods~\cite{ye_duaw_2023,liang_mist_2023,li_preventing_nodate,le_anti-dreambooth_2023} demonstrate that adversarial perturbations can disrupt downstream fine-tuning, yet they typically require iterative per-image optimization at protection time. Such methods can achieve strong disruption, but their computational cost limits their applicability when a large number of generated outputs must be protected before release. Generator-based defenses~\cite{zhu_watermark-embedded_2024} improve efficiency by producing perturbations in a forward pass, but they may provide limited explicit perturbation control, introduce visible artifacts, or suffer from unstable optimization. Therefore, existing methods do not simultaneously provide effective disruption, visual fidelity, explicit budget control, and deployment-level efficiency.

To address this gap, we propose \textbf{WaveGuard}, an output-level, generator-based protection framework for released synthetic images. WaveGuard aims to preserve the perceptual utility of synthetic images for benign viewers while reducing their utility as training data for unauthorized student models. Given a clean generated image and a user-specified perturbation budget, WaveGuard produces a protected image in a single forward pass. Its frequency-aware perturbation generator injects structured, bounded perturbations that remain visually mild in pixel space but alter the latent training signals extracted by downstream substitute models. By enforcing a hard $\ell_\infty$ budget, WaveGuard provides explicit imperceptibility control and supports a practical fidelity--protection--efficiency trade-off for public release scenarios.

Our contributions are threefold. \textbf{First}, we formulate unauthorized distillation from released synthetic images as a practical threat to closed-weight image generative services, and identify output-level protection as a deployable defense setting. \textbf{Second}, we introduce WaveGuard, a single-pass, generator-based protection framework that enforces an explicit $\ell_\infty$ perturbation budget and uses frequency-aware perturbation generation to preserve visual fidelity while disrupting downstream substitute training. \textbf{Third}, through extensive experiments under WikiArt-related unauthorized imitation settings, we show that WaveGuard achieves a favorable efficacy--fidelity--efficiency trade-off compared with representative defenses, and further analyze its transferability and robustness under attacker-side preprocessing.

\section{Related Work}

\subsection{Text-to-Image Generative Models}

Diffusion models~\cite{dhariwal_diffusion_2021,ho_denoising_2020,rombach_high-resolution_2022} have become the dominant framework for text-to-image synthesis, with latent diffusion models (LDMs)~\cite{rombach_high-resolution_2022} enabling efficient high-resolution generation in a compressed latent space. Built on these generative backbones, lightweight adaptation and personalization methods, such as Textual Inversion~\cite{gal_image_2022}, DreamBooth~\cite{ruiz_dreambooth_2023}, LoRA~\cite{hu_lora_2021}, and Custom Diffusion~\cite{kumari_multi-concept_2023}, allow models to rapidly absorb new concepts, identities, and visual patterns from a small number of examples. Furthermore, controllable generation is facilitated by modular mechanisms like ControlNet~\cite{zhang_adding_2023} and T2I-Adapter~\cite{mou_t2i-adapter_2024}, which provide extra guidance via auxiliary pathways. Collectively, these advancements~\cite{park_styleboost_2023, chung_style_2024} allow for the seamless encoding of unique aesthetics and artistic styles into customized models. While these techniques were originally developed for controllable generation and reuse, they also make it easier for an attacker to train a private substitute model from collected synthetic outputs, turning released images into potential distillation data.

\subsection{Adversarial Protection for Diffusion Models}

Recent work studies adversarial protection against downstream personalization and fine-tuning~\cite{shan_glaze_2023,salman_raising_2023,liu_metacloak_2024,hu_who_2025}. A large class of defenses is optimization-based: perturbations are computed per image through iterative procedures that interfere with diffusion denoising or latent representations~\cite{shan_nightshade_2024,liang_adversarial_2023,liang_mist_2023,le_anti-dreambooth_2023,li_preventing_nodate}. These methods often provide strong disruption, but their computational cost can be prohibitive for large-scale output release. Generator-based defenses such as AdvWM~\cite{zhu_watermark-embedded_2024} improve throughput, but practical deployment still requires explicit budget control and strong fidelity guarantees. Our work targets this latter regime: fast output-level protection for released synthetic assets rather than maximum-disruption optimization alone. 

\subsection{Anti-Distillation and Model Protection}

Knowledge distillation is widely used for model compression and transfer, but the same teacher--student learning paradigm can also enable model stealing when teacher outputs are exposed to untrusted learners. Early anti-distillation studies mainly focus on discriminative classifiers, where the defender modifies the supervision signal disclosed to the student, such as soft labels, sparse logits, or adversarially perturbed outputs, so that the teacher remains useful for its intended task while becoming less informative for unauthorized distillation~\cite{ma2021undistillable,ma2022sparse}. More recent work extends this output-centric view to generative models, especially large language models, where generated responses, reasoning traces, or reformulated outputs are adjusted to preserve user-facing utility while reducing their value as distillation data~\cite{savani_antidistillation_2025,li_doge_2025,hartman_hiding_nodate,ding2025information,fang2026towards}. 
Our work brings this output-side anti-distillation perspective to text-to-image generative services, where released synthetic images remain perceptually useful to benign users while becoming less effective as supervision for unauthorized substitute models.

\subsection{Frequency-Aware Perturbation Generation}

Frequency-domain structure plays an important role in both adversarial robustness and perceptual quality. SimAC~\cite{wang_simac_2024} and DDAP~\cite{yang_ddap_2024} highlight the value of frequency-aware perturbation design for diffusion-related attacks. From the generative side, wavelet-based architectures~\cite{yang_wavegan_2022} show that low-frequency structure and high-frequency detail can be disentangled and reused through frequency-specific pathways. Frequency-aware adversarial generation~\cite{zhu_frequency-aware_2023} further indicates that explicit frequency control can reduce visible artifacts. Beyond protection-oriented perturbations, controlled signal design has also been explored from a positive-utility perspective, where learned positive-incentive noise is used to benefit model training~\cite{zhang_variational_pin_2025,zhang_data_augmentation_pin_2026,zhu_beneficial_noise_mllms_2026}. These observations motivate our use of wavelet-based low-frequency residuals and high-frequency skip injection to preserve layout while producing structured perturbations under a strict budget.

\section{Threat Model and Problem Setup}
\label{sec:threat}

\subsection{Substitute Student Training Threat}
We consider a teacher generative model $\mathcal{T}$ deployed in a closed-weight, query-exposed setting. Given a prompt $p \in \mathcal{P}$, the teacher generates an image
\begin{equation}
x \sim \mathcal{T}(p).
\end{equation}

An attacker seeks to steal the capabilities of the teacher model by collecting its outputs for unauthorized knowledge distillation. The attacker collects a dataset of released generated outputs paired with prompts:
\begin{equation}
\mathcal{D} = \{(p_i, x_i)\}_{i=1}^{N},
\end{equation}
where each $x_i$ is a released image produced by the teacher for prompt $p_i$, and $N$ can be large because query-based collection is highly efficient.
The attacker’s goal is to obtain a substitute student model $\mathcal{S}$. Starting from a publicly available pretrained checkpoint, the attacker applies a standard personalization or fine-tuning method $\mathcal{A}$ to the collected dataset:
\begin{equation}
\mathcal{S} = \mathcal{A}(\mathcal{D}).
\end{equation}

The attack succeeds if $\mathcal{S}$ can generate images that match the teacher's distribution 
and reproduce the teacher's generation behavior. In this paper, we instantiate it as \textbf{style knowledge distillation in substitute training}, a concrete and measurable visual form of unauthorized distillation. The core task is therefore to prevent such style mimicry while preserving the utility of released images for benign users.


\subsection{Defense Goals and Constraints}
To protect the teacher outputs before release, we transform each clean image $x$ into a protected image
\begin{equation}
\tilde{x} = \Pi_{\mathrm{adv}}(x).
\end{equation}
The attacker only observes $\tilde{x}$ rather than $x$.

Our defense goals are threefold. \textbf{Utility preservation}: protected outputs should remain visually faithful to the released content. \textbf{Substitute-training disruption}: students trained on protected outputs should replicate the teacher less effectively than students trained on clean outputs. \textbf{Deployment practicality}: protection should be efficient and explicitly budget-controlled enough for large-scale release scenarios. We also consider basic attacker-side preprocessing such as JPEG compression, blur, or purification.

\section{Method}
\label{sec:method}

\begin{figure*}[tb]
  \centering
  \includegraphics[width=0.76\linewidth]{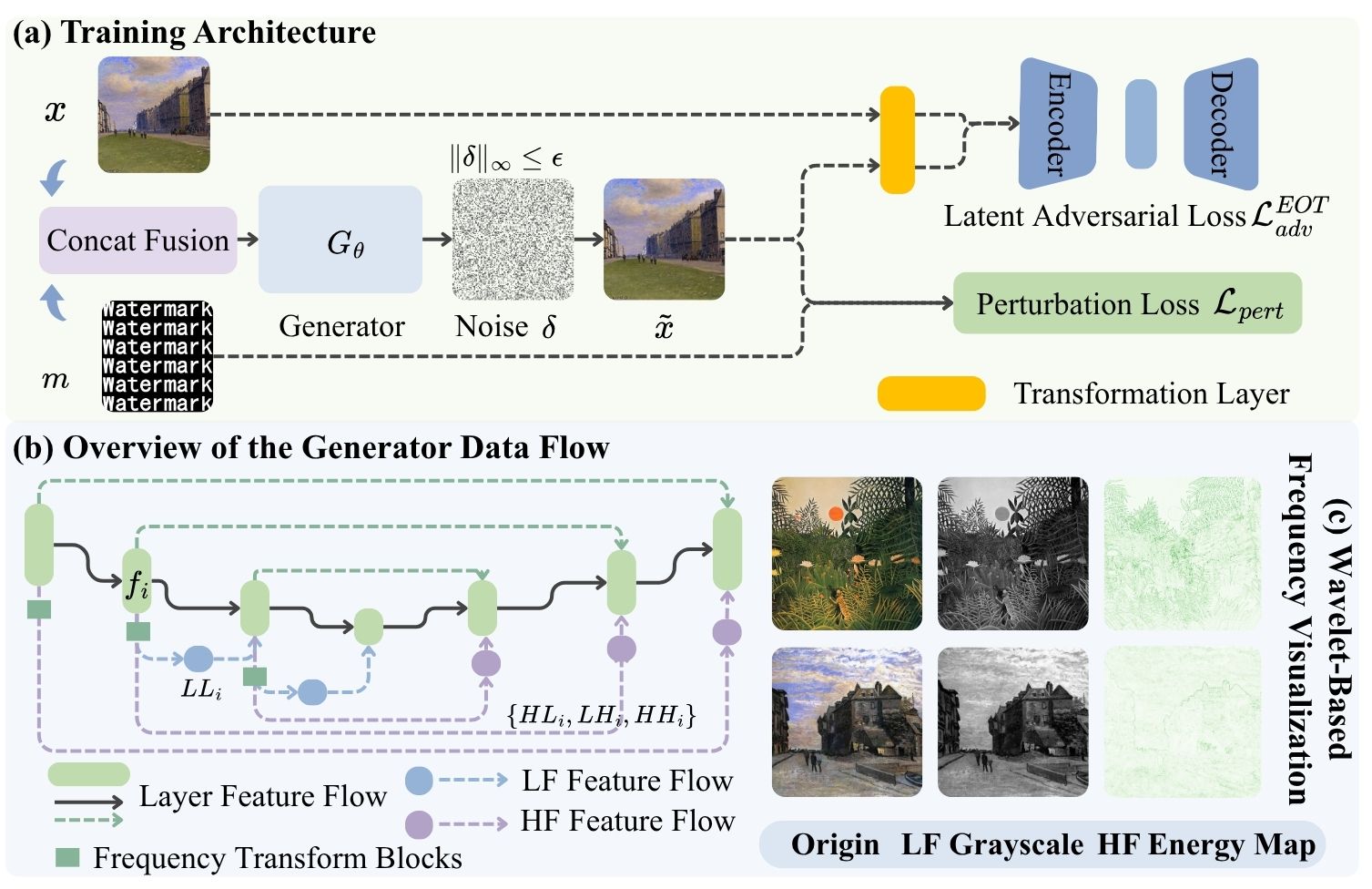}
  \caption{\textbf{Training pipeline and generator data flow.}
  (a) Overview of the WaveGuard training pipeline. The generator predicts bounded adversarial perturbations and adds them to the original image to produce the protected output.
  (b) Data flow of low-frequency (LF) and high-frequency (HF) features in the frequency-aware injection path of the generator.
  (c) Grayscale visualization of the $LL$ subband and high-frequency energy maps of the $\{LH,HL,HH\}$ components.}
  \label{fig:model_structure}
\end{figure*}

\subsection{Overview}
We propose \textbf{WaveGuard}, a generator-based, output-level protection framework against distillation. As illustrated in Figure~\ref{fig:model_structure}(a), WaveGuard employs a \textbf{frequency-aware perturbation generator} $G_\theta(x,m)$ trained with a \textbf{latent adversarial loss} and a \textbf{perturbation loss}. Given a clean image $x$ and a target image $m$, the generator predicts an additive perturbation that is applied to the original image to produce the protected output. At inference time, WaveGuard secures images in a single forward pass, ensuring high efficiency.
We provide details on the generator architecture and optimization objectives in Sections \ref{sec:waveguard_generator} and \ref{sec:training_obj}, respectively.

\subsection{Frequency-aware Perturbation Generator}
\label{sec:waveguard_generator}

We employ a U-Net-style encoder-decoder architecture, $G_\theta$, to generate additive perturbations subject to an explicit magnitude constraint. For a given clean image $x$ and target image $m$, we construct a joint input via channel-wise concatenation:
\begin{equation}
x_{\mathrm{in}} = [x; m].
\end{equation}
The generator, comprising an encoder $\mathrm{Enc}(\cdot)$ and a decoder $\mathrm{Dec}(\cdot)$, maps the combined input to a perturbation $\delta = G_\theta(x_{\mathrm{in}})$. To enhance the visual invisibility of the generated noise, we introduce Frequency Transform Blocks (FTBs). These blocks perform spectral decomposition on latent features, which are then integrated back into the main feature stream through the Frequency-aware Injection Path (FIP). Together, these modules facilitate frequency-aware perturbation generation, ensuring adversarial effectiveness while maintaining low visibility.

\paragraph{\textbf{Frequency Transform Blocks.}}
Haar wavelet transform blocks~\cite{daubechies_wavelet_1990} are employed to decompose features into frequency components. The 1D Haar filters are defined as
\begin{equation}
L^T=\frac{1}{\sqrt{2}}[1,\,1],\qquad H^T=\frac{1}{\sqrt{2}}[-1,\,1],
\end{equation}
where $L$ and $H$ denote the low-pass and high-pass filters, respectively, and $(\cdot)^T$ represents the transpose operator. Specifically, $L$ extracts the local average of the signal, while $H$ captures local differences. In 2D image processing, these filters are applied convolutionally along rows and columns to achieve multi-scale spectral decomposition, inducing four separable 2D analysis kernels. For an intermediate encoder feature map $f_i$, the Frequency Transform Block applies the DWT using these kernels, yielding four subbands:
\begin{equation}
(LL_i,\,LH_i,\,HL_i,\,HH_i)=\mathrm{DWT}(f_i).
\end{equation}
Here $LL_i$ captures low-frequency structures such as global layout and coarse shape, while $\{LH_i,HL_i,HH_i\}$ encode high-frequency details such as edges and contours. Figure~\ref{fig:model_structure}(c) visualizes these components as a low-frequency grayscale map and high-frequency energy maps to characterize their distinct spectral properties. We apply these blocks at multiple scales to obtain multi-level frequency features across layers.

\paragraph{\textbf{Frequency-aware Injection Path.}}
Inspired by frequency-aware generative models~\cite{yang_wavegan_2022}, we leverage multi-level spectral features to enhance the generator's frequency-domain understanding. As illustrated in Figure~\ref{fig:model_structure}(b), WaveGuard strategically injects frequency components into both the encoder and decoder. On the encoder side, since $LL_i$ captures the global appearance and fundamental structural layout~\cite{yang_wavegan_2022}, we employ low-frequency residual connections to preserve global semantics and ensure feature faithfulness during extraction. Specifically, let $f_i$ denote the encoder feature at scale $i$. After wavelet decomposition, we inject $LL_i$ into the subsequent encoding stage as a low-frequency residual to stabilize the feature flow:
\begin{equation}
f_{i+1} = \mathrm{Enc}_{i+1}(f_i) + LL_i.
\end{equation}
Here $\mathrm{Enc}_{i+1}$ denotes the $(i+1)$-th layer of the encoder. Conversely, the high-frequency triplet $H_i=\{LH_i,HL_i,HH_i\}$ is cached as skip information to support fine-grained perturbation synthesis. Previous studies~\cite{zhu_frequency-aware_2023} indicate that significant high-frequency information can be lost during deep feature extraction, often leading to noisy adversarial examples with undesirable aliasing artifacts. To maintain visual fidelity, we explicitly route $H_i$ to the decoder. During decoding, the cached component $H_i$ is fused with the current decoder feature $d_{i+1}$ through the inverse DWT to reconstruct high-fidelity spectral details:
\begin{equation}
\tilde{d}_{i} = \mathrm{IDWT}(d_{i+1},\, H_i).
\end{equation}
Finally, $\tilde{d}_{i}$ is fused with the corresponding spatial skip features from the encoder to balance spectral integrity and spatial precision.

\paragraph{\textbf{Bounded Perturbation Output.}}
The generator outputs a normalized perturbation through a $\tanh$ layer and enforces a hard $\ell_\infty$ bound via explicit scaling:
\begin{align}
\hat{\delta} &= \tanh(\mathrm{Dec}(\mathrm{Enc}(x_{\mathrm{in}}))), \\
\delta &= \epsilon\cdot \hat{\delta}, \\
\tilde{x} &= \Pi_{\mathrm{proj}}(x+\delta),
\end{align}
where $\hat{\delta}$ is the normalized decoder output and $\delta$ is the final bounded perturbation. Here $\epsilon$ denotes the user-specified perturbation budget, while $\Pi_{\mathrm{proj}}(\cdot)$ projects the perturbed sample back to the valid image domain.

\subsection{Training Loss}
\label{sec:training_obj}

\paragraph{\textbf{Latent Adversarial Loss with EOT.}}
Following the targeted adversarial setting~\cite{liang_mist_2023,zhu_watermark-embedded_2024}, we adopt a fixed black-and-white target image $m$ as the optimization target. The objective is to steer the latent representation of the protected image $\tilde{x}$ toward that of $m$, thereby obfuscating the original features extracted from $x$. Specifically, we utilize the frozen VAE encoder $E(\cdot)$ from Stable Diffusion as a surrogate latent encoder. Given a clean image $x$ and the target image $m$, the encoder produces their respective latent embeddings $E(x)$ and $E(m)$. Under this targeted objective, we encourage the protected output to align with the target in the latent space by minimizing
\begin{equation}
\mathcal{L}_{adv} = \mathbb{E}_{x\sim\mathcal{D}} \left[ \|E(\tilde{x})-E(m)\|_2^2 \right].
\end{equation}
This loss compels the generator to produce perturbations that mislead the generative model into capturing features of the target $m$ while disrupting the encoding of the original image semantics.

To enhance robustness against common post-processing operations, we further incorporate the Expectation over Transformations (EOT) framework. Specifically, we apply a set of random differentiable transformations $t\sim\mathrm{T}$ to both $\tilde{x}$ and $m$ during training:
\begin{equation}
\mathcal{L}_{adv}^{\mathrm{EOT}} = \mathbb{E}_{x\sim\mathcal{D}}\, \mathbb{E}_{t\sim\mathrm{T}}
\left[ \|E(t(\tilde{x}))-E(t(m))\|_2^2 \right],
\end{equation}
where $\mathrm{T}$ denotes a distribution of transformations including identity mapping, differentiable JPEG compression, and Gaussian blur.

\paragraph{\textbf{Weighted Perturbation Hinge Loss.}}
To further reduce visual distortion induced by perturbations, we regularize perturbation magnitude using a weighted hinge penalty, as commonly adopted in adversarial training~\cite{zhu_frequency-aware_2023,zhu_watermark-embedded_2024}. Let $\delta=\tilde{x}-x$ denote the additive perturbation. We construct a single-channel template mask $\bar{m}$ from $m$ and define a spatial weight map
\begin{equation}
M = 1 + w\cdot \bar{m},
\end{equation}
where $w$ controls the strength of spatially non-uniform perturbation regularization. The weighted perturbation hinge loss is
\begin{equation}
\mathcal{L}_{pert} =
\mathbb{E}_{x\sim\mathcal{D}}
\left[
\max\left(0,\ \|M\odot\delta\|_2 - c\right)
\right],
\end{equation}
where $c$ controls the penalty threshold.

\paragraph{\textbf{Overall objective.}}
We optimize the generator parameters $\theta$ by minimizing
\begin{equation}
\min_{\theta}\ 
\lambda_{adv}\mathcal{L}_{adv}^{\mathrm{EOT}}
+\lambda_{pert}\mathcal{L}_{pert}.
\end{equation}
Only $G_\theta$ is updated during training, while the target VAE encoder remains frozen.

\section{Experiments}
\label{sec:experiments}

\subsection{Experimental Setup}
\label{sec:exp_setting}
We evaluate WaveGuard under a synthetic-output distillation scenario, where an attacker collects released images from a text-to-image teacher and uses them to train an unauthorized student model. This setting allows us to measure released-image fidelity, substitute-training disruption, and deployment efficiency in a unified evaluation pipeline.
\paragraph{Datasets.}
We use WikiArt~\cite{saleh2015large} (over 80{,}000 artworks; 129 artists, 11 genres, and 27 styles) as the default artistic-style domain. We select 28 representative artists and use 4--6 artworks per artist to build personalized teacher models. For generator training, we sample about 500 additional real artworks from the same artist pool, disjoint from the personalization set. We also build an alternative synthetic training set by generating roughly 18 teacher outputs per artist. Unless otherwise specified, the generator is trained on the real-image set, and the artist split is kept fixed across all experiments.

\paragraph{Teacher and student models.}
Our default teacher is a personalized Stable Diffusion v1.5 (SD1.5) model. The substitute student can be initialized from public open-source latent-diffusion checkpoints, including SD1.5 and SD2.1. Unless otherwise specified, we use SD1.5 for the main student initialization, while the SD2.1 student is discussed in Section~\ref{sec:exp_robust}. We evaluate unauthorized student training with two fine-tuning methods: DreamBooth and Textual Inversion.

\paragraph{Baselines.}
We compare WaveGuard with two iterative baselines, Mist~\cite{liang_mist_2023} and CosAttack~\cite{li_preventing_nodate}, and one generative baseline, AdvWM~\cite{zhu_watermark-embedded_2024}. Unless stated otherwise, all methods are evaluated at $\epsilon=8/255$ under the same $\ell_\infty$ constraint.

\paragraph{Attack protocol.}
Since style mimicry is a concrete manifestation of unauthorized knowledge distillation, we focus on this threat model in our evaluation. The teacher first generates synthetic outputs, which are then protected before release; the attacker collects these protected outputs and trains a substitute model. Concretely, we use a prompt family of 10 variants such as \texttt{``a drawing in the style of sks''} to synthesize teacher outputs for each artist, and the attacker uses those outputs for substitute training.

\paragraph{Metrics.}
For fidelity, we report PSNR, SSIM, and LPIPS between clean images $x$ and protected images $\tilde{x}$; higher PSNR/SSIM and lower LPIPS indicate better visual fidelity. For protection effectiveness, we evaluate style replication between teacher and student outputs using ArtFID~\cite{wright2022artfid} and CSD~\cite{somepalli_measuring_2024}. Lower CSD and higher ArtFID indicate stronger protection, since they reflect weaker style reproduction by the student.

\paragraph{Implementation details.}
All images are generated and evaluated at $512\times512$. WaveGuard trains the perturbation generator for 200 epochs with batch size 8 and Adam optimization, while the target Stable Diffusion v1.5 VAE encoder is frozen. DreamBooth students use a unified fine-tuning setup, and Textual Inversion students follow a consistent 500-step protocol. The same hardware and framework configuration was used across methods for fair runtime and robustness comparisons.

\subsection{Quantitative Results}
\label{sec:exp_main}

Table~\ref{tab:main_results_syn} reports the main results at $\epsilon=8/255$ on this representative visual instantiation, including both DreamBooth and Textual Inversion (TI) students. The key pattern is consistent across metrics: WaveGuard achieves the best visual fidelity among the compared defenses while still providing substantial protection over the no-protection baseline. In particular, Mist obtains the strongest protection on the DreamBooth CSD and ArtFID metrics, whereas WaveGuard attains the best PSNR, SSIM, and LPIPS. For the TI student, WaveGuard remains clearly better than the clean baseline and competitive with CosAttack, although AdvWM and Mist are stronger in this setting. WaveGuard represents a different operating point from maximum-disruption iterative baselines: it prioritizes high-fidelity released outputs and single-pass deployability while maintaining effective protection against substitute training.

\begin{table*}[!t]
\centering
\caption{\textbf{Main results on synthetic-output protection.}
DreamBooth and Textual Inversion (TI) results are reported jointly. WaveGuard achieves the strongest fidelity metrics while maintaining effective protection under both student training routes. Best and second-best values are marked in bold and underlined, respectively.}
\label{tab:main_results_syn}
\small
\setlength{\tabcolsep}{3pt}
\resizebox{0.8\textwidth}{!}{%
\begin{tabular}{l|ccc|cc|cc}
\hline
& \multicolumn{3}{c|}{\textbf{Fidelity}} & \multicolumn{2}{c|}{\textbf{DreamBooth Protection}} & \multicolumn{2}{c}{\textbf{TI Protection}} \\
\textbf{Method} & \textbf{PSNR$\uparrow$} & \textbf{SSIM$\uparrow$} & \textbf{LPIPS$\downarrow$} & \textbf{CSD$\downarrow$} & \textbf{ArtFID$\uparrow$} & \textbf{CSD$\downarrow$} & \textbf{ArtFID$\uparrow$} \\
\hline
No protection & N/A & N/A & N/A & 0.830 & 17.750 & 0.792 & 22.805 \\
Mist~\cite{liang_mist_2023} & 33.079 & 0.878 & 0.144 & \textbf{0.460} & \textbf{30.338} & \underline{0.440} & \underline{31.749} \\
CosAttack~\cite{li_preventing_nodate} & 30.545 & 0.831 & 0.208 & \underline{0.539} & 24.455 & 0.504 & 29.064 \\
AdvWM~\cite{zhu_watermark-embedded_2024} & \underline{34.555} & \underline{0.896} & \underline{0.102} & 0.628 & 23.673 & \textbf{0.439} & \textbf{32.495} \\
WaveGuard (ours) & \textbf{36.201} & \textbf{0.930} & \textbf{0.089} & 0.640 & \underline{24.920} & 0.500 & 30.624 \\
\hline
\end{tabular}
}
\end{table*}

\subsection{Efficiency and Perturbation-Budget Trade-off}
\label{sec:runtime}

\begin{figure}[!t]
\centering
\includegraphics[width=0.8\linewidth]{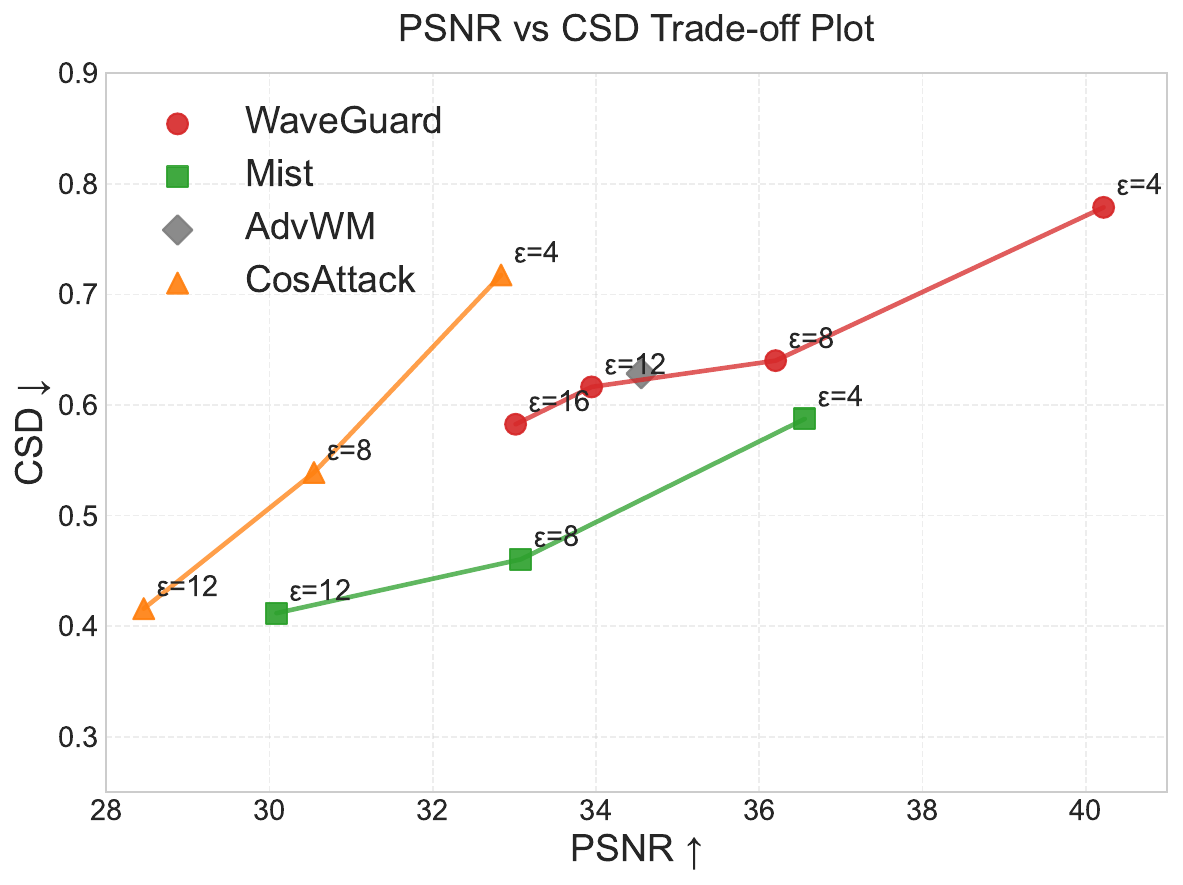}
\caption{\textbf{Budget-controlled fidelity--protection trade-off.}
Each point corresponds to a different perturbation budget.}
\label{fig:runtime_tradeoff}
\end{figure}

Table~\ref{tab:runtime_comparison} summarizes runtime, and Figure~\ref{fig:runtime_tradeoff} shows the PSNR--CSD trade-off under varying perturbation budgets. Generator-based methods are substantially faster than iterative optimization baselines. In particular, WaveGuard and AdvWM operate at similar speed, while WaveGuard attains a more favorable PSNR--CSD trajectory than the iterative baseline CosAttack in the evaluated range. At comparable or better fidelity, WaveGuard provides non-trivial protection gains over no protection and approaches the protection strength of iterative baselines with orders-of-magnitude lower runtime.

\begin{table}[!t]
\centering
\caption{\textbf{Runtime comparison.} Generator-based methods protect images substantially faster than iterative optimization baselines.}
\label{tab:runtime_comparison}
\small
\setlength{\tabcolsep}{3pt}
\resizebox{0.6\columnwidth}{!}{%
\begin{tabular}{lccc}
\hline
\textbf{Method} & \shortstack{\textbf{Raw}\\\textbf{latency (s)}} & \shortstack{\textbf{Batch}\\\textbf{size}} & \shortstack{\textbf{Latency}\\\textbf{(s/image)}} \\
\hline
Mist & 20.41 & 1 & 20.410 \\
CosAttack & 17.72 & 1 & 17.720 \\
AdvWM & 0.11 & 4 & 0.030 \\
WaveGuard & 0.10 & 4 & 0.025 \\
\hline
\end{tabular}
}
\end{table}

\subsection{Qualitative Analysis}
\label{sec:exp_qual}

\begin{figure}[!t]
  \centering
  \includegraphics[width=0.96\linewidth]{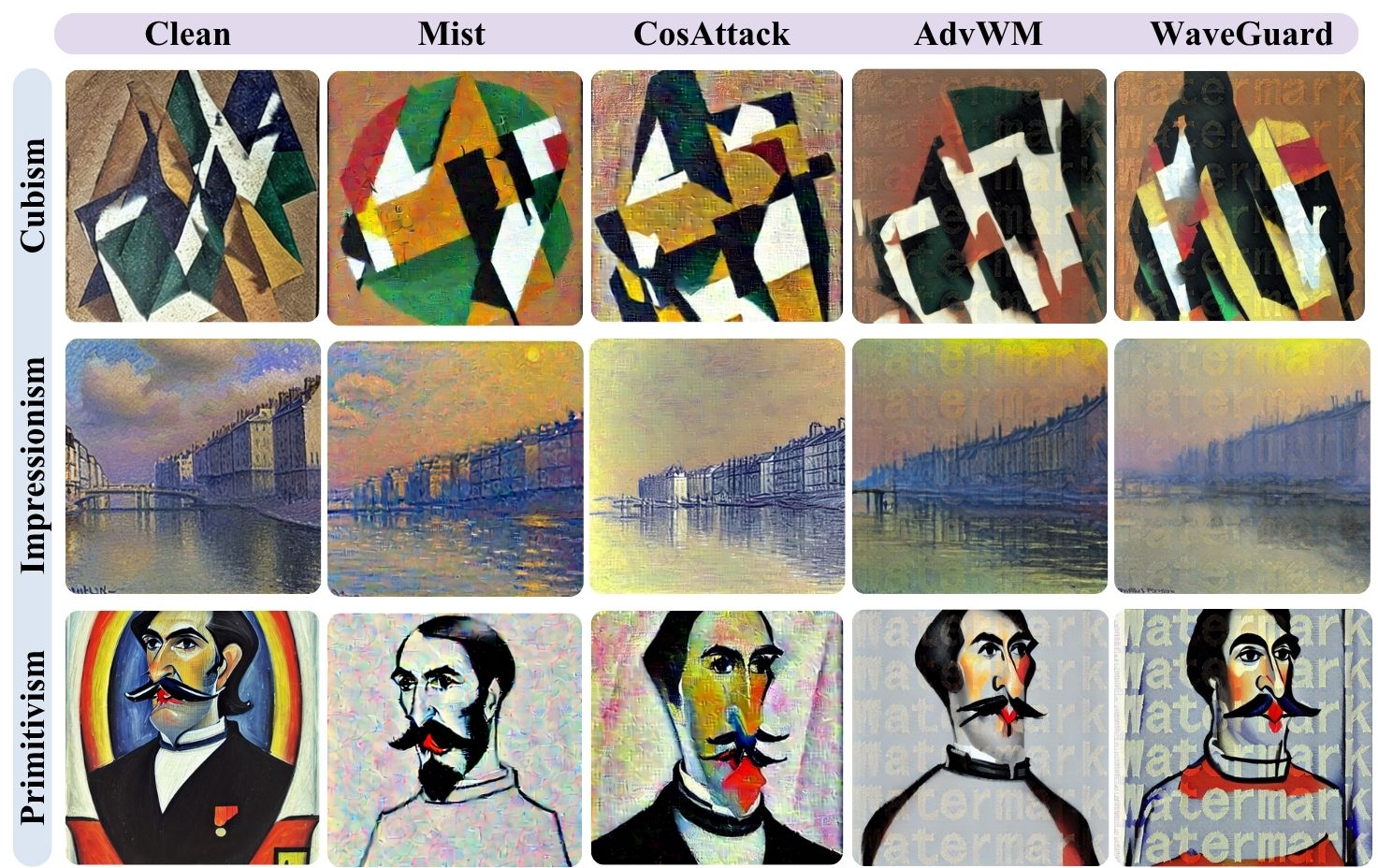}
  \caption{\textbf{Qualitative comparison of DreamBooth student outputs after protection.} Poorer style replication indicates stronger protection. WaveGuard reduces style replication while keeping the released protected images visually closer to the clean images than iterative baselines.}
  \label{fig:atk_res_qualitative}
\end{figure}

\begin{figure}[!t]
  \centering
  \includegraphics[width=0.96\linewidth]{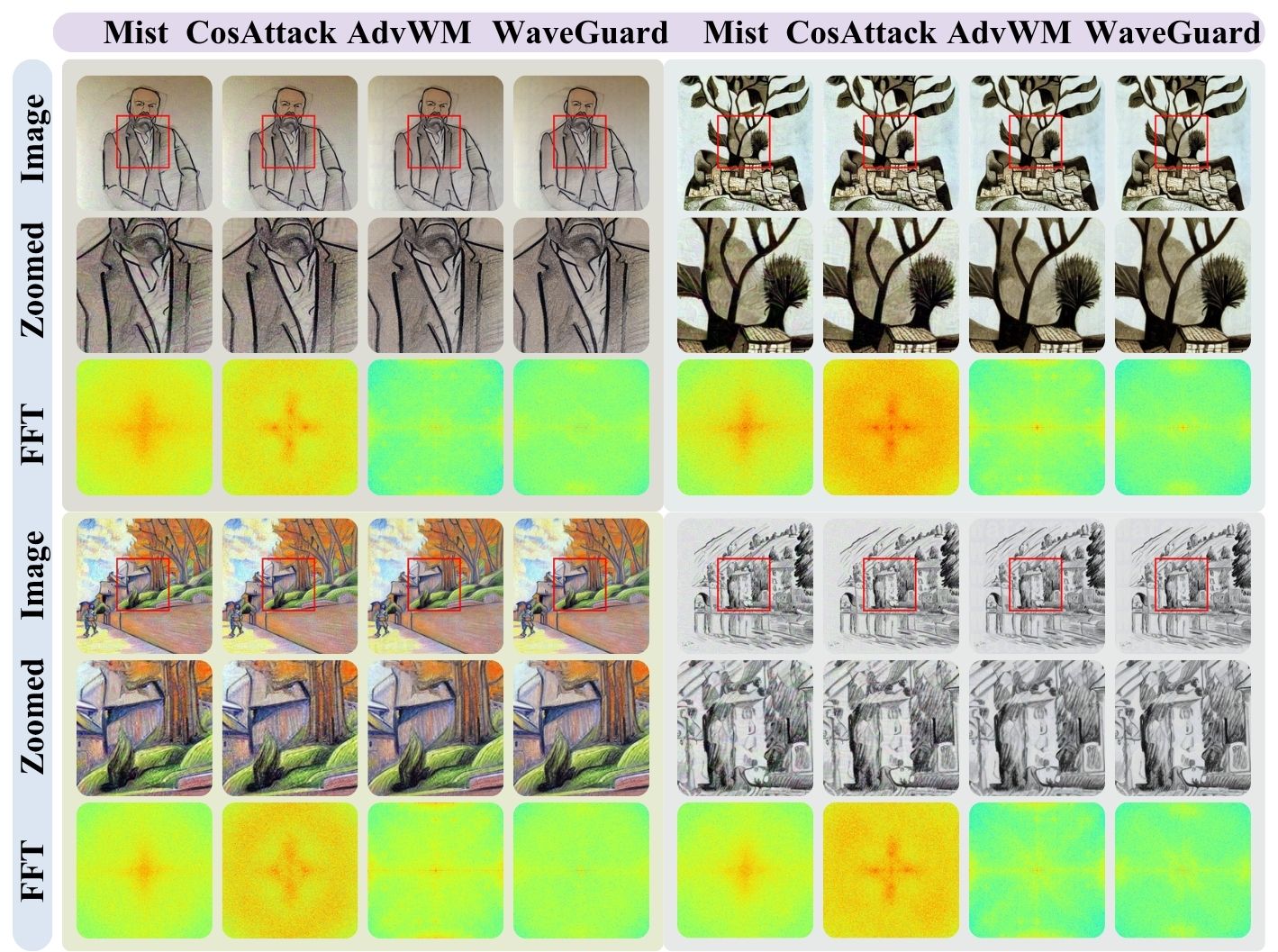}
  \caption{\textbf{Visual and frequency-domain analysis of protected images.} Compared with iterative baselines, WaveGuard shows weaker overall perturbation energy and less visually disruptive frequency spread, which is consistent with its stronger fidelity metrics.}
  \label{fig:fft_vis_comparison}
\end{figure}

Figure~\ref{fig:atk_res_qualitative} evaluates downstream imitation by comparing DreamBooth student outputs trained on images protected by different methods. Figure~\ref{fig:fft_vis_comparison} analyzes the released protected images themselves, including their visual appearance and perturbation spectra. Iterative methods, especially CosAttack, spread perturbation energy more broadly and introduce stronger spatial artifacts. By contrast, WaveGuard produces more structured perturbations with weaker visible distortion, consistent with its role as a high-fidelity defense point rather than a maximum-disruption method.

\subsection{Ablation Study}
\label{sec:exp_ablation}

We further analyze the design of our frequency module in Table~\ref{tab:module_ablation_full}. The full model achieves the best PSNR, SSIM, CSD, and ArtFID among these variants. Removing LF injection slightly reduces fidelity and weakens protection, suggesting that low-frequency feature routing contributes to the overall trade-off. Removing HF injection leaves fidelity nearly unchanged but weakens protection more noticeably, indicating that high-frequency skip routing helps produce perturbations that remain effective for disrupting substitute training. 
Beyond the frequency-aware architecture, we further study two design choices in the supplementary material: the target image used in the latent objective and the data source used to train the perturbation generator. 

\begin{table}[!t]
\centering
\caption{\textbf{Module-wise differential ablation under the same training protocol.} The full WaveGuard model provides the strongest overall fidelity--protection balance.}
\label{tab:module_ablation_full}
\small
\setlength{\tabcolsep}{3pt}
\resizebox{\columnwidth}{!}{%
\begin{tabular}{lccccc}
\hline
\textbf{Method} & \textbf{PSNR$\uparrow$} & \textbf{SSIM$\uparrow$} & \textbf{LPIPS$\downarrow$} & \textbf{CSD$\downarrow$} & \textbf{ArtFID$\uparrow$} \\
\hline
WaveGuard (full) & \textbf{36.201} & \textbf{0.930} & 0.089 & \textbf{0.640} & \textbf{24.920} \\
w/o LF & 35.806 & 0.925 & 0.089 & 0.694 & 24.165 \\
w/o HF & 36.193 & 0.929 & \textbf{0.088} & 0.704 & 23.048 \\
w/o Wave module & 35.996 & 0.927 & 0.089 & 0.706 & 23.460 \\
\hline
\end{tabular}
}
\end{table}

\subsection{Robustness to Attacker Strategies}
\label{sec:exp_robust}

We next evaluate practical attacker-side strategies that may weaken protection before substitute training. 

\paragraph{Transfer to another latent-diffusion checkpoint.}
Table~\ref{tab:robust_transfer} evaluates a student initialized from Stable Diffusion v2.1. Although WaveGuard is trained with a Stable Diffusion v1.5 VAE surrogate, it still degrades style replication for the SD2.1 student, indicating cross-checkpoint transferability.

\paragraph{Mixed clean/protected training.}
We also evaluate mixed-data training, where attackers combine clean and protected samples under a fixed budget of 10 samples per artist.
Table~\ref{tab:mixed_training} reports the results. Increasing the fraction of protected data consistently improves disruption. A 50\% protected ratio nearly matches the all-protected setting in CSD, although a gap remains in ArtFID.

\begin{table}[!t]
\centering
\caption{\textbf{Transfer to another latent-diffusion checkpoint (SD2.1).}}
\label{tab:robust_transfer}
\small
\setlength{\tabcolsep}{4pt}
\resizebox{0.66\columnwidth}{!}{%
\begin{tabular}{c|cc}
\hline
\textbf{Training data} & \textbf{CSD$\downarrow$} & \textbf{ArtFID$\uparrow$} \\
\hline
No protection & 0.845 & 20.311 \\
WaveGuard ($\epsilon=8/255$) & 0.663 & 25.657 \\
WaveGuard ($\epsilon=12/255$) & \textbf{0.582} & \textbf{27.648} \\
\hline
\end{tabular}
}
\end{table}

\begin{table}[!t]
\centering
\caption{\textbf{Mixed clean/protected training.}}
\label{tab:mixed_training}
\small
\setlength{\tabcolsep}{6pt}
\resizebox{0.66\columnwidth}{!}{%
\begin{tabular}{c|cc}
\hline
\textbf{\# clean / 10} & \textbf{CSD$\downarrow$} & \textbf{ArtFID$\uparrow$} \\
\hline
10 (all clean) & 0.844 & 18.716 \\
8 & 0.816 & 19.621 \\
5 & 0.664 & 24.255 \\
0 (all protected) & \textbf{0.663} & \textbf{26.687} \\
\hline
\end{tabular}
}
\end{table}

\paragraph{Preprocessing robustness.}

We evaluate JPEG compression, Gaussian blur, and DiffPure~\cite{nie_diffusion_2022} as attacker-side preprocessing.
As shown in Table~\ref{tab:robust_preproc}, all three preprocessing operations weaken protection. WaveGuard remains clearly stronger than no protection under JPEG and blur, whereas DiffPure reduces the protection effect much more substantially and brings the metrics closer to the no-protection baseline.

\begin{table}[!t]
\centering
\caption{\textbf{Preprocessing robustness of WaveGuard.}}
\label{tab:robust_preproc}
\small
\setlength{\tabcolsep}{5pt}
\begin{tabular}{l|cc}
\hline
\textbf{Preprocessing} & \textbf{CSD$\downarrow$} & \textbf{ArtFID$\uparrow$} \\
\hline
No protection & 0.859 & 16.951 \\
WaveGuard (clean) & \textbf{0.613} & \textbf{27.790} \\
+ JPEG & 0.760 & 20.816 \\
+ Blur & 0.688 & 23.314 \\
+ DiffPure & 0.825 & 19.327 \\
\hline
\end{tabular}
\end{table}


\section{Conclusion}
We studied the problem of safeguarding generative models against unauthorized knowledge distillation through output-level adversarial protection. We presented WaveGuard, a frequency-aware, single-pass perturbation generator that protects images released by text-to-image generative services. Rather than pursuing maximum disruption alone, WaveGuard targets a high-fidelity deployment regime: it enforces an explicit perturbation budget, preserves visual quality, and provides efficient protection at release time. In the studied WikiArt-style distillation scenario, WaveGuard achieves the best fidelity among compared defenses while retaining meaningful protection and strong runtime advantages. Further analyses characterize transferability and vulnerabilities under alternative students, models, targets, training mixes, and preprocessing. We also find that strong purification such as DiffPure can substantially weaken the protection. Future work will extend the evaluation to broader domains and improve robustness against stronger adaptive preprocessing. Overall, these results suggest that output-level, generator-based protection is a promising direction for safeguarding generative models against unauthorized knowledge distillation.

\bibliography{ref}

\clearpage 
\appendix

\section{Experimental Information}
\label{sec:repro}

\subsection{Hardware and Software Environment}
\label{sec:env}
We use two execution environments, one for student fine-tuning/evaluation and one for WaveGuard generator training, as summarized in Table~\ref{tab:env_versions}.

\begin{table}[!htbp]
\centering
\caption{\textbf{Hardware and framework versions used in experiments.}}
\label{tab:env_versions}
\setlength{\tabcolsep}{4pt}
\begin{tabular}{lc}
\hline
\textbf{Item} & \textbf{Configuration} \\
\hline
Student Train/Eval & RTX 4080 SUPER \\
WaveGuard Train  & RTX 4090 D \\
PyTorch & 2.3.1+cu121 \\
diffusers & 0.29.2 \\
accelerate & 1.12.0 \\
\hline
\end{tabular}
\end{table}

\subsection{Dataset Splits}
\label{sec:artist_list}
Our default training and evaluation set includes the following 28 artists
(the 8 artists used for supplementary sampling are highlighted in yellow):
\begin{center}
\colorbox{gray!15}{%
\parbox{0.94\linewidth}{%
\centering\small\ttfamily
alberto-magnelli, amedeo-modigliani, andrei-ryabushkin,\par
\colorbox{yellow!35}{caspar-david-friedrich},\par
\colorbox{yellow!35}{claude-monet}, edvard-munch, el-greco,\par
\colorbox{yellow!35}{giuseppe-arcimboldo}, gustav-klimt,\par
\colorbox{yellow!35}{gustave-dore}, henri-rousseau,\par
jackson-pollock, jean-francois-millet, lucian-freud,\par
\colorbox{yellow!35}{lyubov-popova}, marcel-duchamp,\par
maria-helena-vieira-da-silva, mark-rothko, max-beckmann,\par
\colorbox{yellow!35}{mikhail-nesterov},\par
\colorbox{yellow!35}{mikhail-vrubel}, odilon-redon, oswaldo-guayasamin,\par
pablo-picasso, \colorbox{yellow!35}{paul-gauguin}, paul-signac,\par
tivadar-kosztka-csontvary, willem-de-kooning
}%
}
\end{center}

\subsection{Training Details}
\label{sec:training_details}

\subsubsection{Generator Training Details}

The implementation of WaveGuard uses a frozen Stable Diffusion v1.5 VAE encoder as a feature-space target and trains only the perturbation generator.
Unless otherwise specified, training uses $200$ epochs, batch size $8$, Adam optimizer with learning rate $1\times10^{-3}$, and random seed $42$. All images are resized to $512 \times 512$ with center crop. The perturbation budget is set to $\epsilon=8/255$ in $[0,1]$ space (equivalently $2 \times 8/255$ in $[-1,1]$ space).
During training, we apply expectation-over-transformation (EOT) augmentations: differentiable JPEG simulation with quality sampled from $[30,95]$ and Gaussian blur with $\sigma\sim U(0.1,1.5)$, each applied independently with probability $0.5$.
For JPEG/blur robustness evaluation, we apply JPEG compression with quality factor $50$ and Gaussian blur with kernel size $5$ to the generated protected images.

\subsubsection{Teacher/Student Training Details}
\label{sec:student_details}

Our teacher-side personalization is based on DreamBooth fine-tuning with Stable Diffusion v1.5. On the student side, we evaluate two training routes, DreamBooth and Textual Inversion, initialized from Stable Diffusion v1.5 or v2.1. We next describe the parameter settings for each route.

\textbf{DreamBooth Settings.}
All DreamBooth student fine-tuning experiments follow a unified protocol.
We train DreamBooth at resolution $512\times512$ with batch size $1$, gradient accumulation steps $1$, and $400$ optimization steps. Optimization uses 8-bit Adam with a constant learning rate of $5\times 10^{-6}$, with no warmup steps. For each artist folder in the protected training set, we train one personalized
DreamBooth student using the instance prompt ``a painting in sks style''.
Prior preservation is enabled with the class prompt ``a painting in art style'', with the number of class images set to $100$, and a prior loss weight of $1.0$.

\textbf{Textual Inversion Settings.}
For Textual Inversion (TI) evaluation, we train one TI embedding per artist using each method's corresponding protected training set, while keeping the TI hyperparameters identical across methods. TI training uses the initializer token ``art'', resolution $512$, batch size $1$, gradient accumulation steps $4$, learning rate $5\times10^{-4}$ (constant schedule, no warmup), and $500$ optimization steps. We assign a unique placeholder token to each artist and evaluate the checkpoint at step $500$.

\subsection{Prompt Set}
\label{sec:prompt_stats}
For DreamBooth, we use the token ``sks''. For Textual Inversion, we use a learned placeholder token such as ``ti\_vango''.
\begin{center}
\colorbox{gray!15}{%
\parbox{0.94\linewidth}{%
\small\ttfamily
"a painting in sks style",\par
"a painting in the style of sks",\par
"an artwork in sks style",\par
"an artwork in the style of sks",\par
"a picture in sks style",\par
"a picture in the style of sks",\par
"a drawing in sks style",\par
"a drawing in the style of sks",\par
"a masterpiece in sks style",\par
"a masterpiece in the style of sks"
}%
}
\end{center}
\section{Architecture and Implementation}
\label{sec:model}

\subsection{WaveGuard Architecture}
\label{sec:layer_table}

WaveGuard uses a U-Net-style generator with wavelet-based skip connections.
Given a clean image $x\in[0,1]^{3\times H\times W}$ and a watermark target
$m\in[0,1]^{C\times H\times W}$, we map $m$ to three channels when $C=1$ and
concatenate it with $x$:
$\hat{x}=\mathrm{concat}(x,m)\in\mathbb{R}^{6\times H\times W}$. The main encoder branch is downsampled by stride-2 convolutions, while fixed Haar
wavelet decomposition provides LL features for encoder fusion and HF features for
decoder skip injection.
Starting from a bottleneck feature at $H/16\times W/16$, the decoder reconstructs
a full-resolution perturbation map through wavelet unpooling, projection, and refinement.
The output is a normalized perturbation map $\Delta_{\mathrm{norm}}\in[-1,1]^{3\times H\times W}$,
which is scaled by the perturbation budget $\epsilon$.
Table~\ref{tab:waveguard_arch} summarizes the layer-by-layer architecture.
In the table, WP/WP2 denote two wavelet pooling variants, WUP denotes wavelet unpooling,
HF$_k$/LL$_k$ denote the high-/low-frequency branches at stage $k$, and ``skip'',
``proj'', and ``ref'' denote skip fusion, channel projection, and feature refinement,
respectively.

\begin{table}[!t]
\centering
\scriptsize
\caption{\textbf{WaveGuard Architecture.}}
\label{tab:waveguard_arch}
\setlength{\tabcolsep}{3pt}
\begin{tabular}{lllll}
\hline
\textbf{Stage} & \textbf{Op} & \textbf{Channels} & \textbf{Scale} & \textbf{Wavelet} \\
\hline
Input      & concat$(x,m)$         & $3{+}3\rightarrow6$   & $H\times W$           & concat fusion \\
Enc-1      & Conv5,s1 + WP         & $6\rightarrow32$      & $H\rightarrow H$      & keep HF$_1$ \\
Enc-2      & Conv3,s2 + WP         & $32\rightarrow64$     & $H\rightarrow H/2$    & keep HF$_2$ \\
Enc-3      & Conv3,s2 + WP2        & $64\rightarrow128$    & $H/2\rightarrow H/4$  & +LL$_1$, keep HF$_3$ \\
Enc-4      & Conv3,s2 + WP2        & $128\rightarrow128$   & $H/4\rightarrow H/8$  & +LL$_2$, keep HF$_4$ \\
\hline
Bottleneck & Conv3,s2              & $128\rightarrow128$   & $H/8\rightarrow H/16$ & +LL$_3$ \\
\hline
Dec-4      & WUP + skip + ref      & $128\rightarrow128$   & $H/16\rightarrow H/8$ & inject HF$_4$ \\
Dec-3      & WUP + skip + ref      & $128\rightarrow128$   & $H/8\rightarrow H/4$  & inject HF$_3$ \\
Dec-2      & WUP + proj + ref      & $128\rightarrow64$    & $H/4\rightarrow H/2$  & inject HF$_2$ \\
Dec-1      & WUP + proj + ref      & $64\rightarrow32$     & $H/2\rightarrow H$    & inject HF$_1$ \\
Out        & Conv5,s1 + Tanh       & $32\rightarrow3$      & $H$                   & $\Delta_{\mathrm{norm}}\in[-1,1]$ \\
\hline
\end{tabular}
\end{table}

The perturbation is computed as $\Delta=\epsilon\cdot\Delta_{\mathrm{norm}}$, and the final protected image is
$x_{\mathrm{adv}}=\mathrm{clip}(x+\Delta,0,1)$.
Unless otherwise specified, we use $\epsilon=8/255$.

\section{Additional Experimental Results and Analyses}
\label{sec:supp_additional_results}

We include a qualitative illustration of substitute training, a real-image protection comparison for consistency with prior evaluation settings, additional target/source design analyses, and an EOT ablation for attacker-side preprocessing.

\subsection{Qualitative Illustration of Substitute Training}
\begin{figure}[tb]
  \centering
  \includegraphics[width=\linewidth]{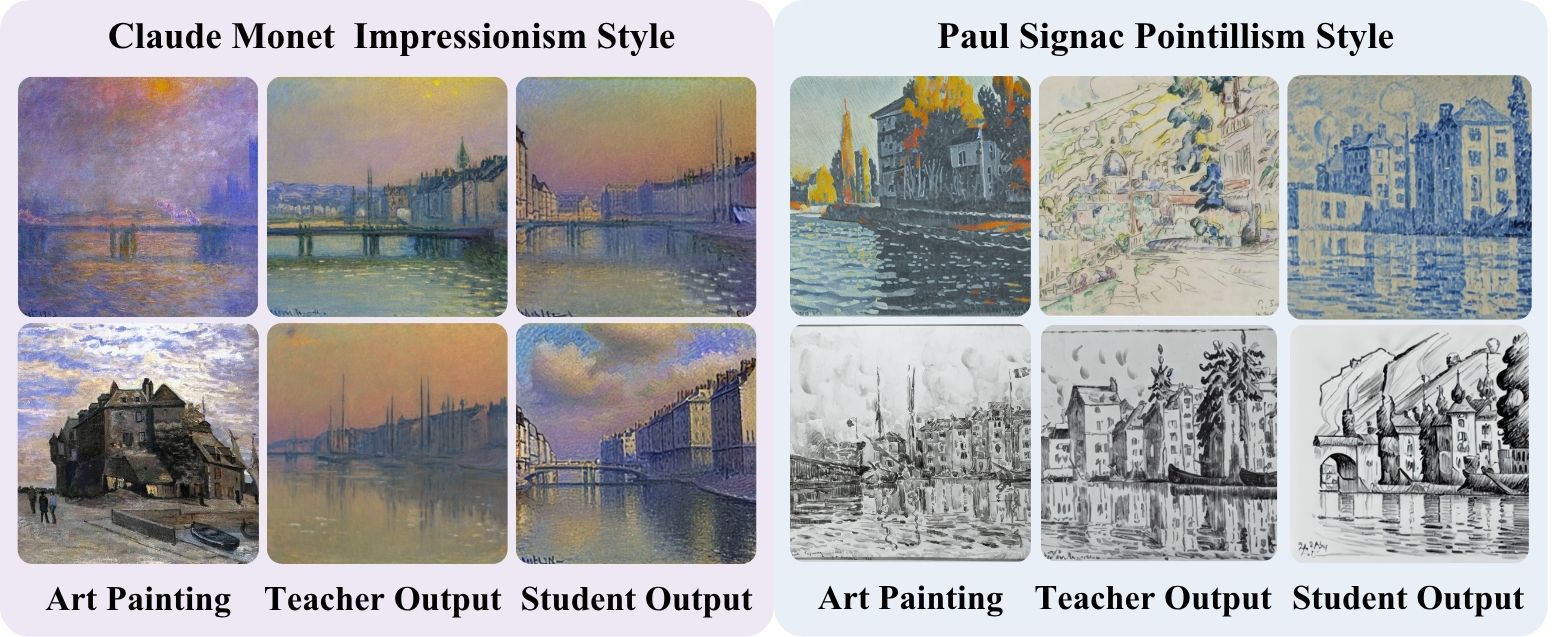}
  \caption{\textbf{Qualitative illustration of student training.} A student initialized from a public checkpoint can be fine-tuned on teacher outputs to imitate the teacher's style.}
  \label{fig:threat_example_supp}
\end{figure}

Figure~\ref{fig:threat_example_supp} illustrates the substitute-training threat considered in this work. 
A student initialized from a public checkpoint can be fine-tuned on teacher-generated outputs to mimic the teacher's style.

\subsection{Real-image Protection Task}

Although our main experiments focus on protecting synthetic images released by a generative service, we also evaluate WaveGuard under a real-image protection setting to facilitate comparison with prior work. In this setting, real WikiArt images are perturbed before release, and the attacker trains a downstream student model on the perturbed images. Table~\ref{tab:real_image_protection} reports the results. WaveGuard achieves the best imperceptibility metrics and the highest ArtFID, while maintaining competitive CSD, suggesting that the proposed generator also provides a favorable fidelity--protection trade-off beyond synthetic outputs.

\begin{table}[t]
\centering
\caption{\textbf{Real-image protection results.} WaveGuard achieves the best imperceptibility and the highest ArtFID, while maintaining competitive CSD.}
\label{tab:real_image_protection}
\small
\setlength{\tabcolsep}{4pt}
\resizebox{\columnwidth}{!}{%
\begin{tabular}{l|ccc|cc}
\hline
& \multicolumn{3}{c|}{\textbf{Imperceptibility}} & \multicolumn{2}{c}{\textbf{Protection}} \\
\textbf{Method} & \textbf{PSNR$\uparrow$} & \textbf{SSIM$\uparrow$} & \textbf{LPIPS$\downarrow$} & \textbf{CSD$\downarrow$} & \textbf{ArtFID$\uparrow$} \\
\hline
No protection & N/A & N/A & N/A & 0.743 & 26.546 \\
Mist~\cite{liang_mist_2023} & 33.077 & 0.878 & 0.144 & \underline{0.450} & 31.631 \\
CosAttack~\cite{li_preventing_nodate} & 30.191 & 0.819 & 0.227 & \textbf{0.418} & 30.051 \\
AdvWM~\cite{zhu_watermark-embedded_2024} & \underline{34.232} & \underline{0.888} & \underline{0.118} & 0.485 & \underline{31.648} \\
\hline
\rowcolor{green!6} WaveGuard (ours) & \textbf{36.593} & \textbf{0.931} & \textbf{0.088} & 0.510 & \textbf{32.879} \\
\hline
\end{tabular}%
}
\end{table}

\subsection{Target and Training Source Analyses}
\label{sec:exp_target_source}

Beyond the frequency-aware architecture, we analyze two design choices that may affect the learned protection behavior: the target image used in the latent objective and the data source used to train the perturbation generator.

\paragraph{Target-image design.}

\begin{figure}[!t]
\centering
\includegraphics[width=0.98\linewidth]{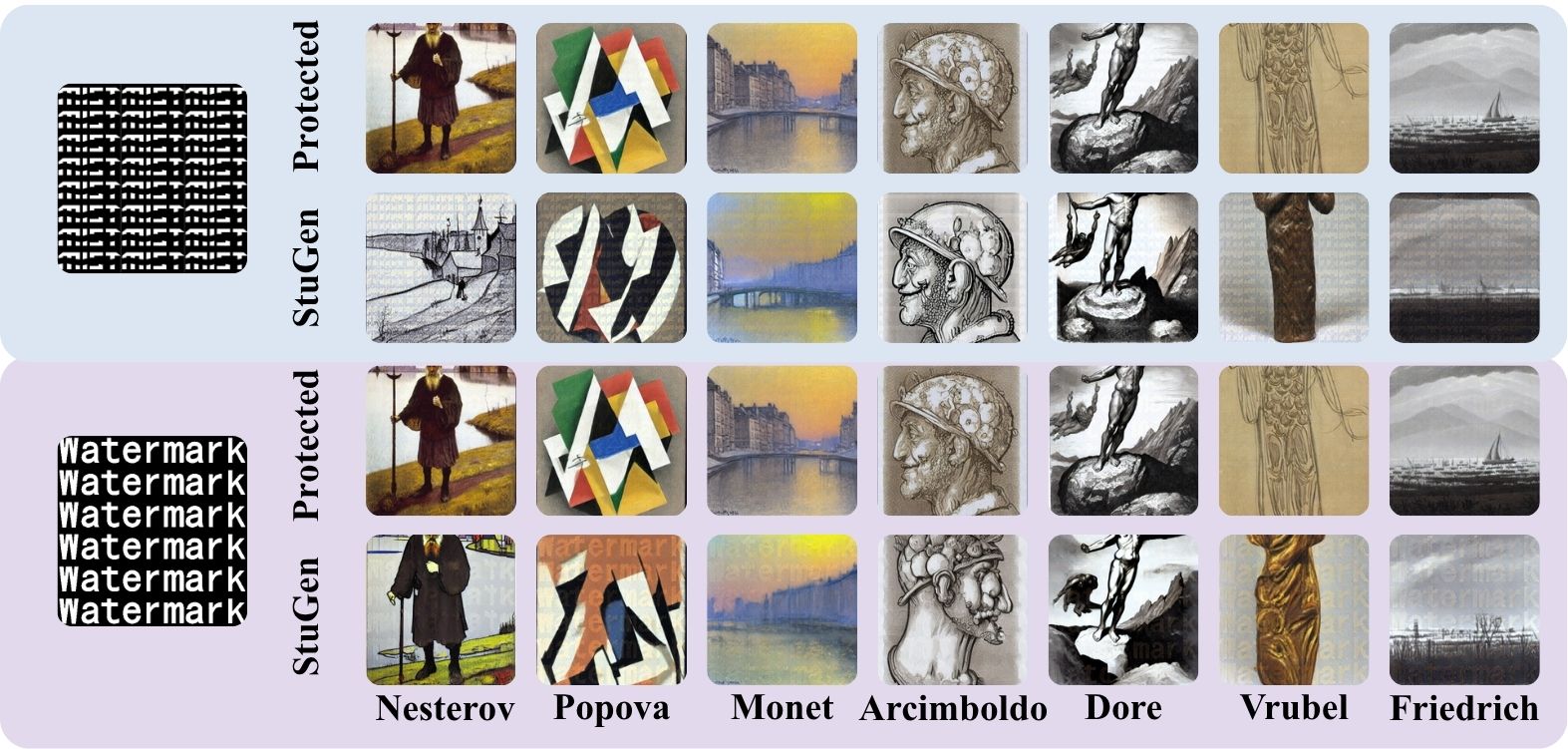}
\caption{\textbf{Effect of different target-image designs.} The top row uses a Mist-style target image, and the bottom row uses the default watermark target image.}
\label{fig:diff_wm_supp}
\end{figure}

Table~\ref{tab:wm_target_quant_supp} and Figure~\ref{fig:diff_wm_supp} compare the default binary target~\cite{hu_who_2025} with a higher-contrast Mist-style target~\cite{liang_mist_2023}. The Mist-style target produces somewhat stronger protection, while the binary target provides better fidelity. This comparison supports the view that target design itself controls an additional fidelity--protection trade-off. This result also indicates that the target image should be treated as a tunable design choice rather than a fixed universal optimum.

\begin{table}[!t]
\centering
\caption{\textbf{Quantitative comparison under different target images.} The target design controls an additional trade-off between image fidelity and protection strength.}
\label{tab:wm_target_quant_supp}
\small
\setlength{\tabcolsep}{3pt}
\resizebox{\columnwidth}{!}{%
\begin{tabular}{lccccc}
\hline
\textbf{Target} & \textbf{PSNR$\uparrow$} & \textbf{SSIM$\uparrow$} & \textbf{LPIPS$\downarrow$} & \textbf{CSD$\downarrow$} & \textbf{ArtFID$\uparrow$} \\
\hline
Watermark image & \textbf{36.201} & \textbf{0.930} & \textbf{0.089} & 0.640 & 24.920 \\
Mist-style image & 35.451 & 0.921 & 0.091 & \textbf{0.590} & \textbf{25.068} \\
\hline
\end{tabular}
}
\end{table}

\paragraph{Generator training source.}
For training the protection generator, we compare two data sources in the synthetic-output protection setting: $\mathcal{G}_{\text{real}}$, trained on real WikiArt images, and $\mathcal{G}_{\text{syn}}$, trained on teacher-generated outputs. As shown in Table~\ref{tab:train_source_syn_supp}, training on real WikiArt images achieves better overall protection, especially on CSD and ArtFID, while the synthetic source only slightly improves LPIPS. 

\begin{table}[!t]
\centering
\caption{\textbf{Training source comparison in synthetic-output protection.}
Positive $\Delta_{\text{syn-better}}$ indicates that training on teacher-generated synthetic outputs outperforms training on real images.}
\label{tab:train_source_syn_supp}
\small
\setlength{\tabcolsep}{4.5pt}
\renewcommand{\arraystretch}{1.12}
\begin{tabular}{c|ccc|cc}
\hline
\textbf{Source} & \textbf{PSNR$\uparrow$} & \textbf{SSIM$\uparrow$} & \textbf{LPIPS$\downarrow$} & \textbf{CSD$\downarrow$} & \textbf{ArtFID$\uparrow$} \\
\hline
$\mathcal{G}_{\text{real}}$ & \textbf{36.201} & \textbf{0.930} & 0.089 & \textbf{0.640} & \textbf{24.920} \\
$\mathcal{G}_{\text{syn}}$  & 35.974 & 0.927 & \textbf{0.086} & 0.727 & 23.474 \\
\rowcolor{green!6} $\Delta_{\text{syn-better}}$ & -0.227 & -0.003 & +0.003 & -0.087 & -1.446 \\
\hline
\end{tabular}
\end{table}

\subsection{EOT Analysis under Preprocessing Attacks}
\label{sec:supp_eot_preproc}
The main paper reports preprocessing robustness under JPEG compression, Gaussian blur, and DiffPure. Here, we further isolate the effect of EOT during WaveGuard training. Table~\ref{tab:eot_preproc_supp} reports the metric degradation after preprocessing. EOT improves robustness to blur and provides a small CSD improvement under JPEG. Its effect on JPEG ArtFID is negligible, indicating that EOT helps most when the preprocessing operation matches the differentiable augmentations used during generator training.

\begin{table}[!t]
\centering
\caption{\textbf{Impact of EOT on preprocessing.} Lower $\Delta_{CSD}$ and lower $\Delta_{ArtFID}$ indicate smaller degradation after preprocessing.}
\label{tab:eot_preproc_supp}
\small
\setlength{\tabcolsep}{5pt}
\begin{tabular}{l|l|cc}
\hline
\textbf{EOT} & \textbf{Attack} & \textbf{$\Delta_{CSD}\downarrow$} & \textbf{$\Delta_{ArtFID}\downarrow$} \\
\hline
w/ & Blur & \textbf{0.075} & \textbf{4.476} \\
w/o & Blur & 0.161 & 4.934 \\
w/ & JPEG & \textbf{0.147} & 6.974 \\
w/o & JPEG & 0.149 & \textbf{6.881} \\
\hline
\end{tabular}
\end{table}


\end{document}